\documentclass[floatfix,twocolumn,nofootinbib,superscriptaddress,a4paper,preprintnumbers]{revtex4-1}

\usepackage{amsmath}
\usepackage{amssymb}
\usepackage{graphicx}
\usepackage{ulem}
\usepackage[mathscr]{eucal}
\usepackage{color}
\usepackage[T1]{fontenc} 
\usepackage{slashed}
\usepackage{xspace}
\usepackage{hyperref}
\usepackage{xspace}

\newcommand\Vevacious{{\tt Vevacious}\xspace}

\newcommand\SARAH{{\tt SARAH}\xspace}
\newcommand\SPheno{{\tt SPheno}\xspace}

\begin{document}

\preprint{KA-TP-21-2017}

\title{Reopen parameter regions in Two-Higgs Doublet Models}

\author{Florian Staub}
\affiliation{Institute for Theoretical Physics (ITP), Karlsruhe Institute of Technology, Engesserstra{\ss}e 7, D-76128 Karlsruhe, Germany}
\affiliation{Institute for Nuclear Physics (IKP), Karlsruhe Institute of Technology, Hermann-von-Helmholtz-Platz 1, D-76344 Eggenstein-Leopoldshafen, Germany}


\begin{abstract}
The stability of the electroweak potential is a very important constraint for models of new physics. At the moment, it is standard for Two-Higgs doublet models (THDM), singlet or triplet extensions of the standard model to perform these checks at tree-level. However, these models are often studied in the presence of very large couplings. Therefore, it can be expected 
that radiative corrections to the potential are important. We study these effects at the example of the THDM type-II and find that loop corrections can revive more than 50\% of the phenomenological viable points which are ruled out by the tree-level vacuum stability checks. Similar effects are expected for other extension of the standard model. 
\end{abstract}
\maketitle

\section{Introduction}
The discovery of a scalar boson  at the Large Hadron Collider with a mass of around 125~GeV was a milestone for particle physics \cite{Aad:2012tfa,Chatrchyan:2012xdj}. This state has all expected properties of the long searched-for Higgs boson, and all particles predicted by the standard model of particle physics (SM) have finally been found. Moreover, the measured mass itself lies in a particular interesting range: combining this information with the one of the measured top mass $m_t$, one finds that the scalar potential of the SM becomes unstable at very high energies \cite{Degrassi:2012ry}. This is not a fundamental problem for the SM. The lifetime of the vacuum we are living in exceeds the age of the universe by many orders of magnitude because of the large separation of the two minima. \\
As soon as extensions of the SM with more scalars are considered, new vacua much 'closer' to ours can appear. Thus, it is necessary to check which combinations of parameters in these models provide a stable or at least sufficiently long-lived potential with correct electroweak symmetry breaking (EWSB). In supersymmetric models it was already realised in the 80s that dangerous charge and colour breaking minima can occur in specific directions of the scalar potential \cite{Nilles:1982dy, AlvarezGaume:1983gj, Derendinger:1983bz, Claudson:1983et, Kounnas:1983td, Drees:1985ie, Gunion:1987qv, Komatsu:1988mt, Langacker:1994bc,Casas:1995pd, Casas:1996de}. In the recent years, these constrains were proven to be even too weak. Other dangerous minima were discovered with numerical methods \cite{CamargoMolina:2012hv,Camargo-Molina:2013sta,Chowdhury:2013dka,Blinov:2013fta,MOLINA:2014uha,Chattopadhyay:2014gfa,Camargo-Molina:2014pwa,Hollik:2016dcm,Beuria:2016cdk,Dreiner:2016wwk,Krauss:2017nlh} and the impact of loop and thermal corrections was analysed \cite{CamargoMolina:2012hv,Camargo-Molina:2013sta}. In contrast, the vacuum stability of non-supersymmetric models is still mainly checked at tree-level. For instance, the tree-level potential of two-Higgs doublet models (THDM) has been studied intensively in literature \cite{Klimenko:1984qx,Velhinho:1994np,Ferreira:2004yd,Barroso:2005sm,Maniatis:2006fs,Ivanov:2006yq,Ivanov:2007de,Ivanov:2008er,Ginzburg:2009dp}, and very compact conditions for the stability of the electroweak (ew) potential were found. These results where also generalised to other non-supersymmetric model \cite{Ivanov:2010ww,Ivanov:2010wz,Robens:2015gla,Muhlleitner:2016mzt}. However, it is often not checked how robust these conditions are against radiative corrections. \\
It is known from the minimal supersymmetric standard model (MSSM) that radiative corrections can have an impact on the vacuum stability, but often the conclusion 'stable' or 'unstable' doesn't change once a suitable renormalisation scale is chosen \cite{Camargo-Molina:2013sta}. The reason is that in the MSSM all couplings in the scalar potential are $O(g^2)$, i.e. moderately small. This must not be the case in THDMs: since often, masses and not couplings are chosen as input, in principle, any size of couplings can appear. Usually, the tree-level perturbativity constraints \cite{Kanemura:1993hm,Horejsi:2005da} are applied which filter out points with very large couplings $\gg 4\pi$. Nevertheless, quartic couplings $O(10)$ are not rare. Thus, large loop effects due to these huge couplings aren't surprising at all. As we will see, for a large fraction of points these corrections stabilise the potential. Only in a few cases they destabilise it. This is similar to what has been observed in a singlet extension and the inert THDM, see Refs.~\cite{Chen:2017qcz,Swiezewska:2015paa,Ferreira:2015pfi} \\
This letter is organised as follows: in sec.~\ref{sec:model} the chosen conventions for the THDM are summarised and the used methods to check vacuum stability at the tree- and loop-level are explained. In sec.~\ref{sec:numerics} the numerical setup is presented and the overall impact of the loop corrections is discussed. We summarise in sec.~\ref{sec:conclusion}

\section{THDM and Vacuum stability}
\label{sec:model}
\begin{figure}[tb]
\includegraphics[width=\linewidth]{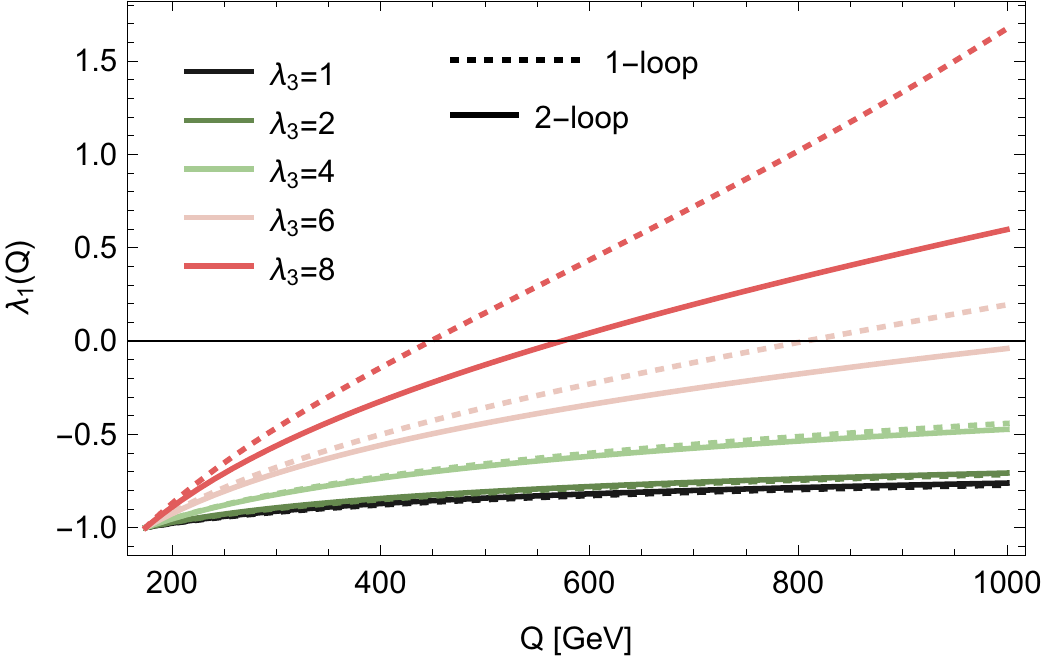}
\caption{Toy example for the running of $\lambda_1$ including only $Y_t=1.1$, $g_3=1.15$ and different values of $\lambda_3$. The dashed lines show the running at one-loop and the full lines at two-loop.}
\label{fig:RGE}
\end{figure}

\begin{figure}[tb]
\includegraphics[width=\linewidth]{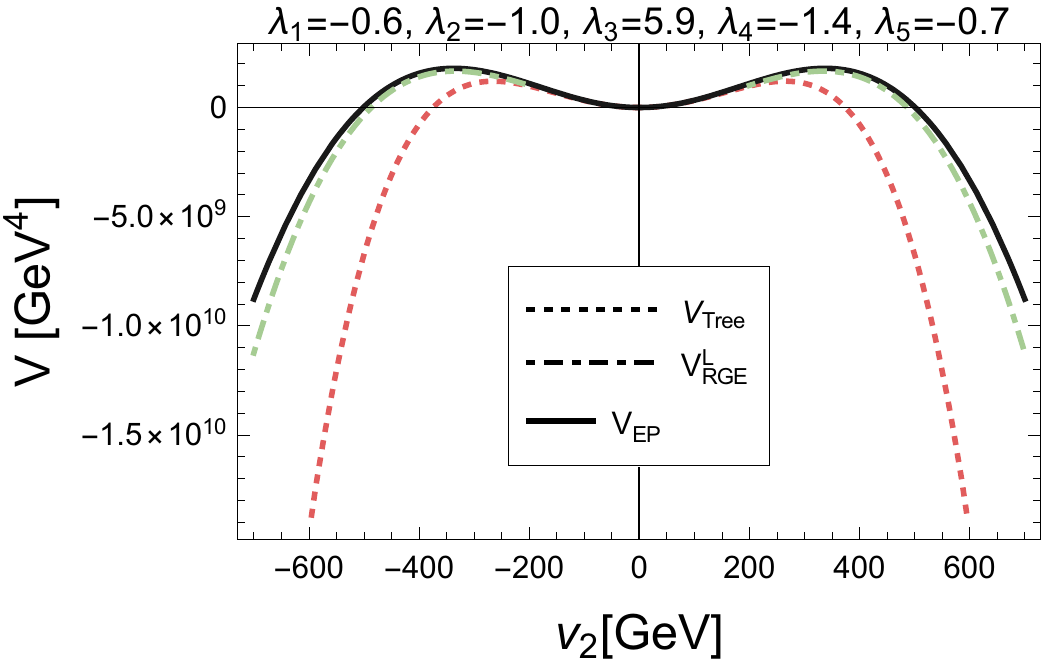} \\[5mm]
\includegraphics[width=\linewidth]{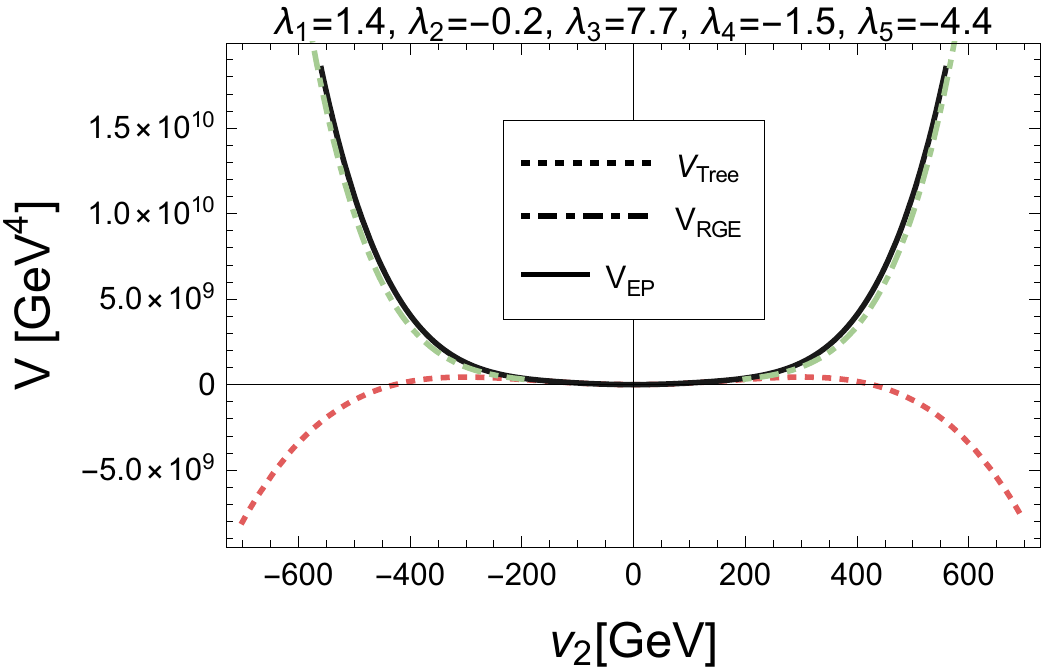}
\caption{The scalar potential for $v_1=0$ for 
two points which are 
unbounded from below at tree-level. We compare here the tree-level potential (dashed red) with the RGE improved potential using two-loop running
(green dot-dashed) and the one-loop effective potential (full black). }
\label{fig:CompUFB}
\end{figure}

\begin{figure}
\includegraphics[width=\linewidth]{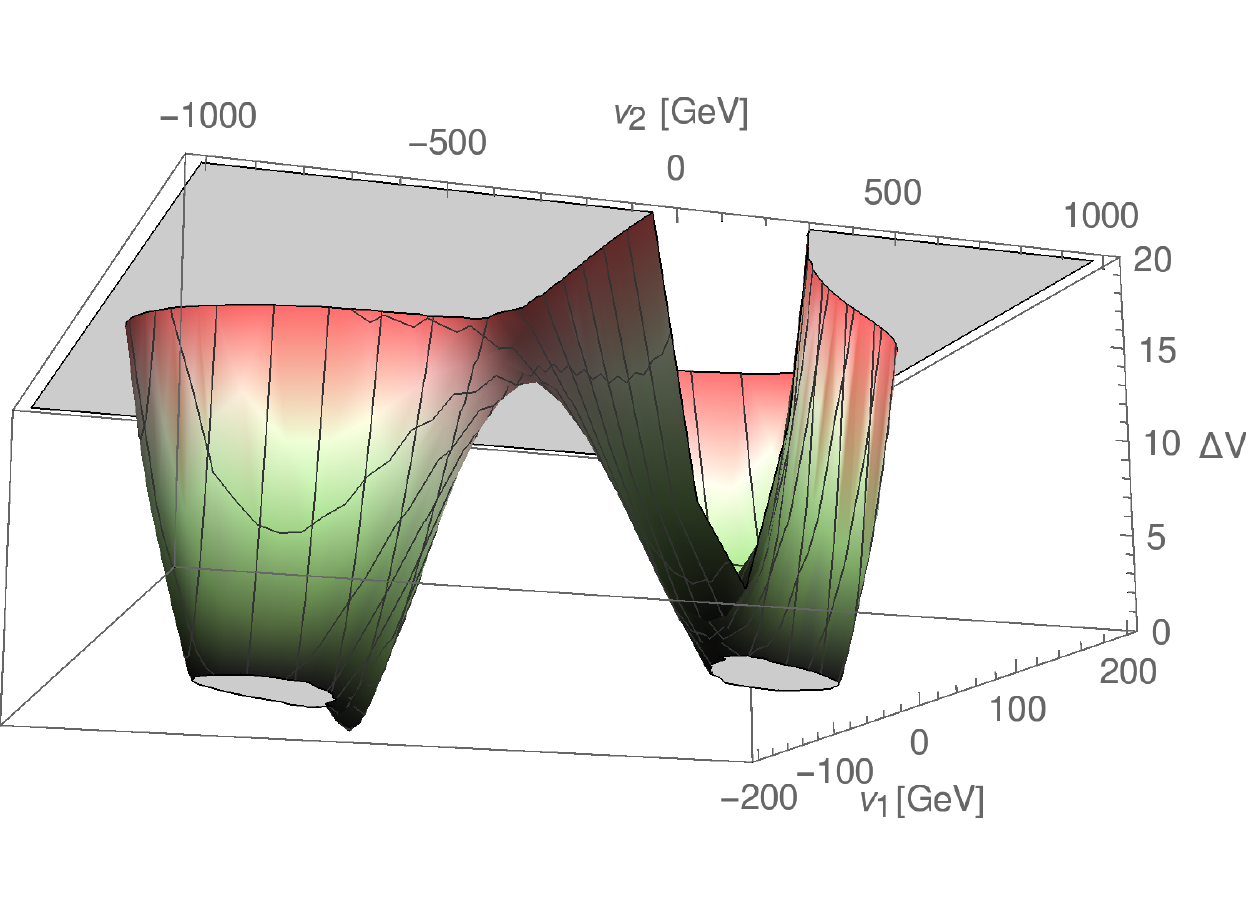} \\
\includegraphics[width=\linewidth]{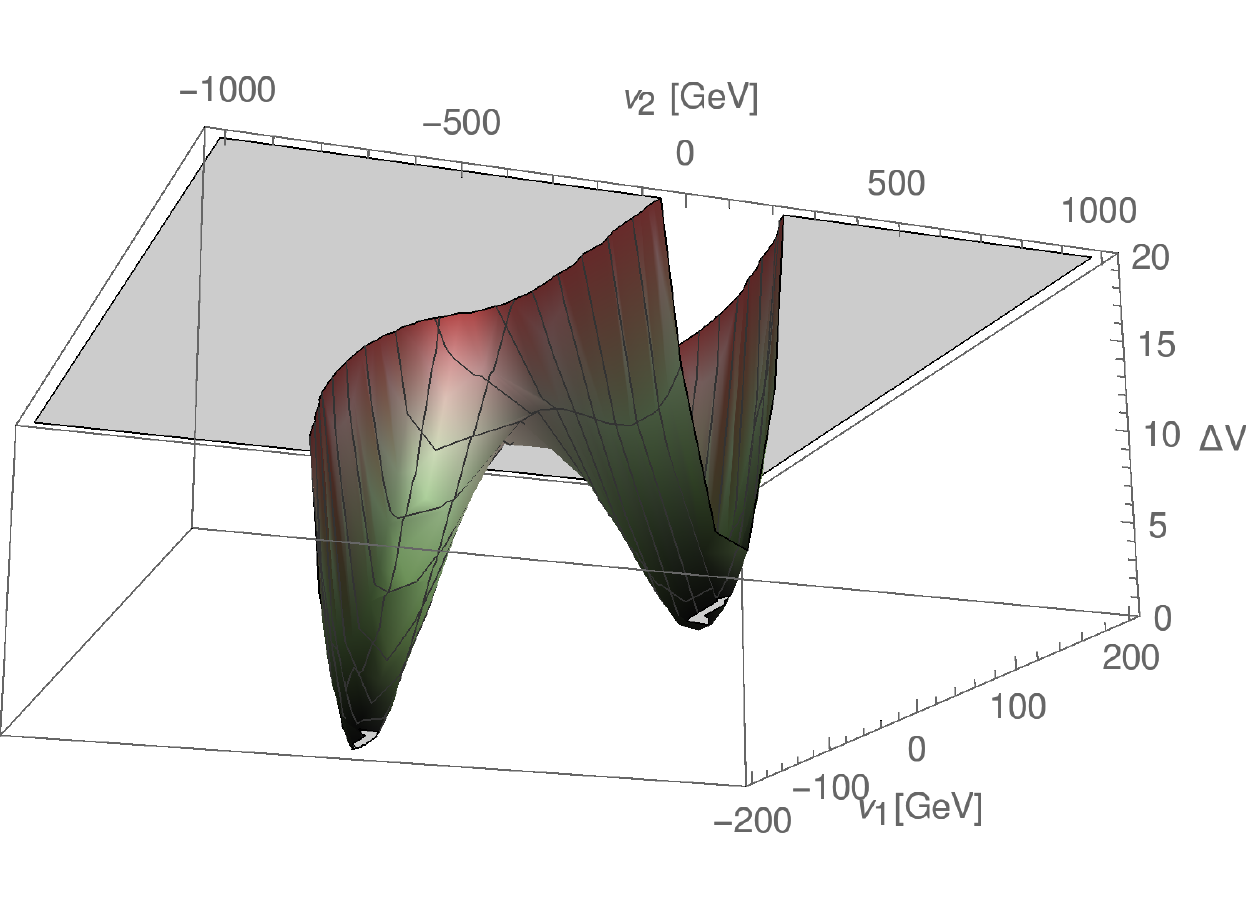} 
\caption{The scalar potential at tree- (top) and one-loop (bottom) for the parameter point
$\lambda_1 = 2.4$, $\lambda_2=0.03$, $\lambda_3=9.8$, $\lambda_4=-4.1$, $\lambda_5=0.7$, $m_{12}=-0.81~\text{TeV}^2$.
Shown is $\Delta V = [V(v_1,v_2)-V(v_1^{ew},v_2^{ew})] \times 10^{-8}$ in units of $\text{GeV}^4$ with the correct ew VEVs $v_1^{ew}=160.1$~GeV, $v_2^{ew}=184.3$~GeV.
}
\label{fig:3Dmeta}
\end{figure}

The scalar potential of a CP conserving THDM  with softly broken $Z_2$ symmetry reads\footnote{We are using in the following the conventions of the model 
as implemented in \SARAH\cite{Staub:2008uz,Staub:2009bi,Staub:2010jh,Staub:2012pb,Staub:2013tta,Staub:2015kfa}. Moreover, we use the arrangement of the Yukawa couplings of type-II. Since the main effects come from the scalar sector itself, 
the results are expected  hardly to change for other variants of THDMs.}
\begin{align}
V_{\rm Tree} = & \lambda_1 |H_1|^4 + \lambda_2 |H_2|^4 + \lambda_3 |H_1|^2 |H_2|^2 + \lambda_4 |H^\dagger_2 H_1|^2 \nonumber \\ 
& \label{eq:pot} \hspace{-1cm} + m_1^2 |H_1|^2 + m_2^2 |H_2|^2   +  \left(m_{12} H_1^\dagger H_2 + \frac12 \lambda_5 (H_2^\dagger H_1)^2 + \text{h.c.}\right)
\end{align}
After EWSB, the neutral components of the two Higgs states receive Vacuum expectation values (VEVs) as
\begin{equation}
H_i = \left(\begin{array}{c} H_i^+ \\ \frac{1}{\sqrt{2}}\left(\phi_i + i \sigma_i + v_i \right) \end{array} \right) \quad i=1,2
\end{equation}
with $\sqrt{v_1^2+v_2^2} = v\simeq 246$~GeV and $\tan\beta=\frac{v_2}{v_1}$. The mass spectrum consists of superposition of these gauge eigenstates, i.e.
$(\phi_1, \phi_2) \to (h,H)$, $(\sigma_1, \sigma_2) \to (G,A)$ and $(H^+_1, H^+_2) \to (G^+,H^+)$. 
Here, $G$ and $G^+$ are the Goldstone modes of the $Z$ and $W$ boson. The mixing in these sectors is fixed by $\tan\beta$, while in the CP-even sector a rotation angle $\alpha$ defines the transition from gauge to mass eigenstates. In practical applications, one can trade the physical masses $m_h$, $m_H$, $m_A$ and $m_{H^+}$ as well as $\tan\alpha$ for the quartic couplings. The necessary relations are
\begin{align}
\label{eq:l1}
\lambda_1 = & \frac{1 + t_\beta^2}{2 (1 + t_\alpha^2) v^2} \left(m_H^2 + m_{12} t_\beta + t_\alpha^2 (m_h^2 + m_{12} t_\beta) \right) \\
 \lambda_2  = & \frac{1 + t_\beta^2}{2 (1 + t_\alpha^2) t_\beta^3 v^2} \left(m_{12} + m_{12} t_\alpha^2 + t_\beta (m_h^2  + m_H^2 t_\alpha^2 ) \right)  \\
 \lambda_3  = & \frac{1}{(1 + t_\alpha^2) t_\beta v^2} \Big[m_h^2 t_\alpha + 2 m_{H^+}^2 (1 + t_\alpha^2) t_\beta \nonumber \\
  & + m_h^2 t_\alpha t_\beta^2 - m_H^2 t_\alpha (1 + t_\beta^2) + m_{12} (1 + t_\alpha^2) (1 + t_\beta^2)\Big] \\
 \lambda_4  = &\frac{1}{t_\beta v^2}\left(-m_{12} + m_A^2 t_\beta - 2 m_{H^+}^2 t_\beta - m_{12} t_\beta^2 \right) \\
 \label{eq:l5}
\lambda_5  = & \frac{1}{t_\beta v^2}\left(-m_{12} - m_A^2 t_\beta - m_{12} t_\beta^2 \right)
\end{align}
with $t_\beta = \tan\beta$ and $t_\alpha = \tan\alpha$. This has the advantage that physical observables instead of Lagrangian parameters can be chosen as input. However, one needs
to be careful since a randomly chosen set of masses could easily correspond to a problematic set of quartic couplings: for very large couplings perturbativity will be spoilt and also unitarity can be violated. Therefore, the first constraints which are usually  applied are those for tree-level unitarity which, roughly spoken,  remove points where combinations of $\lambda$'s are larger than $8 \pi$. The next set of theoretical constraints are those for a stable vacuum. The tree-level conditions to prevent unbounded from below (UFB) directions in the potential are
\cite{Deshpande:1977rw}
\begin{eqnarray}
&\lambda_1 > 0 , \quad \lambda_2 > 0 ,\quad \lambda_3 + 2 \sqrt{{\lambda_1}{\lambda_2}} > 0 & \\
&\lambda_3 + \lambda_4 - |\lambda_5| + 2 \sqrt{{\lambda_1}{\lambda_2}} > 0 &
\end{eqnarray}
while the condition to have no deeper vacua than the ew one is \cite{Barroso:2013awa}
\begin{eqnarray}
\label{eq:CheckMeta}
-m_{12} \left(m_1^2 - \sqrt{\frac{\lambda_1}{\lambda_2}} m_2^2 \right) \left[t_\beta - \left(\frac{\lambda_1}{\lambda_2}\right)^{1/4}\right] > 0 
\end{eqnarray}
These conditions involve the tree-level quartic couplings which are calculated from the chosen masses and angles. However, it is well known from the SM that for large field extension the tree-level potential gets unreliable. In this case one should consider the renormalisation group equation (RGE) improved potential where the parameters are replaced by their running, i.e. scale dependent, values. The running of the quartic coupling in the SM is dominated by the contributions from the top quark which let it run negative at very high scales. 
In the THDM, the one-loop $\beta$-functions for $\lambda_1$ and $\lambda_2$ are given by
\begin{align}
\beta_{\lambda_1}^{(1)} & =  
24 \lambda_{1}^{2} +2 \lambda_3( \lambda_{3} +  \lambda_4) +\lambda_{4}^{2}+\lambda_{5}^{2}+\dots \\ 
\beta_{\lambda_2}^{(1)} & =  
24 \lambda_{2}^{2} +2 \lambda_3( \lambda_{3} +  \lambda_4 )+\lambda_{4}^{2}+\lambda_{5}^{2} \nonumber \\ 
 & \hspace{1cm} +12 \lambda_2 Y_t^2-6 Y_t^4 +\dots 
\end{align}
where the dots indicate subdominant contributions involving $g_1$, $g_2$. Thus, for large $\lambda_{3,4,5}$ the slope of the running will change, i.e.  $\lambda_{1,2}$ increase with the energy scale. To exemplify this, we show in Fig.~\ref{fig:RGE} the scale dependence of $\lambda_1$ in a  toy example involving only $\lambda_1$, $\lambda_3$, $Y_t$ and $g_3$. 
When starting with $\lambda_1=-1$, the coupling becomes already positive below 1~TeV for $\lambda_3> 6$. This points towards a stabilisation of the potential at not too high energies. Since the scale at which $\lambda_1$ changes its sign is not far from the ew scale, an one-loop fixed order calculation can be expected to catch the dominant effects. Therefore, we will consider in the following also the one-loop effective potential
\begin{equation}
V_{EP}^{(1)} = V_{\rm Tree} + V^{(1)}_{\rm CT}  + V^{(1)}_{\rm CW} 
\end{equation}
Here,  $V^{(1)}_{\rm CT}$ is the counter-term (CT) potential which is discussed below. The Coleman-Weinberg 
potential $V^{(1)}_{\rm CW}$ is given by \cite{Coleman:1973jx}
\begin{equation}
V^{(1)}_{\rm CW} = \frac{1}{16\pi^2} \sum_{i}^{\text{all fields}} r_i s_i C_i m_i^4 \left(\log\frac{m_i^2}{Q^2} - c_i \right)
\end{equation}
with $r_i = 1$ for real bosons, otherwise 2; $C_i =3$ for quark, otherwise 1; $\{s_i,c_i\}=\{-\frac{1}{2},\frac{3}{2}\}$ for fermions, $\{\frac{1}{4},\frac{3}{2}\}$ for scalars 
and $\{\frac{3}{4},\frac{5}{6}\}$ for vector bosons. \\
The CT potential is calculated from $V_{\rm Tree}$ with all parameters $x$ replaced by $x+\delta x$. $\delta x$ are the CTs which are usually chosen to cancel all loop corrections to the masses and angles, i.e. the input values are the on-shell ones. One can derive a suitable set of CTs from the renormalisation conditions
\begin{align}
T_i^{CT} + t_i & \equiv 0\\
M^{2,\rm CT}_{ij} + \delta_{ij} \frac{t_i}{v_i} - \Pi_{ij}& \equiv 0
\end{align}
Here, $T_i^{CT}$ and $M^{2,\rm CT}$ are the first and second derivative of the CT potential, and $t_i$ and $\Pi$ are the loop corrections to the one- and two-point functions. 
The crucial point is that the derived 
CTs depend on the ew VEVs, i.e. they give a cancellation between $V_{\rm CT}$  and $V_{\rm CW}$ {\it only} at the ew minimum, but not at other positions of the potential. \\
Having all the machinery at hand, we can compare now the results for the tree-level, RGE improved\footnote{We use the full two-loop RGEs as calculated with \SARAH based on the generic results of Refs.~\cite{Machacek:1983tz,Machacek:1983fi,Machacek:1984zw}} and the one-loop effective potential. This is done in Fig.~\ref{fig:CompUFB} for 
two points which suffer 
from UFB directions at tree-level in the direction $v_1=0,v_2\to \infty$.
We see that the loop corrections have as expected a clear impact on the shape of the potential. 
In the first example, the value of $\lambda_2$ is -1 and the other quartic couplings are not large enough to stabilise the potential in the direction of $v_2$. In contrast, in the second 
example with $\lambda_2 =-0.2$  the UFB direction disappears at the loop-level and the point becomes absolutely stable. Of course, also the situation is possible that the UFB direction disappears at the loop level, but new minima appear which are deeper than the ew one. 	\\
Similarly, we find that the tree-level check for deeper minima than the ew one, eq.~(\ref{eq:CheckMeta}), can lead to a wrong conclusion about the stability of a point. 
We show at one example in Fig.~\ref{fig:3Dmeta} how significantly the shape of the scalar potential can change when going to the loop level. 
We see that the two global minima, which are at tree-level 25\% deeper than the ew one, 
have completely disappeared at the one-loop level. Similarly, one can find also the opposite: points which look stable at the tree-level become metastable at the loop-level. It is now interesting to see how big the fraction of points is where the conclusion about the stability changes at the loop-level.

\section{Results}
\label{sec:numerics}
\begin{figure}[tb]
\includegraphics[width=\linewidth]{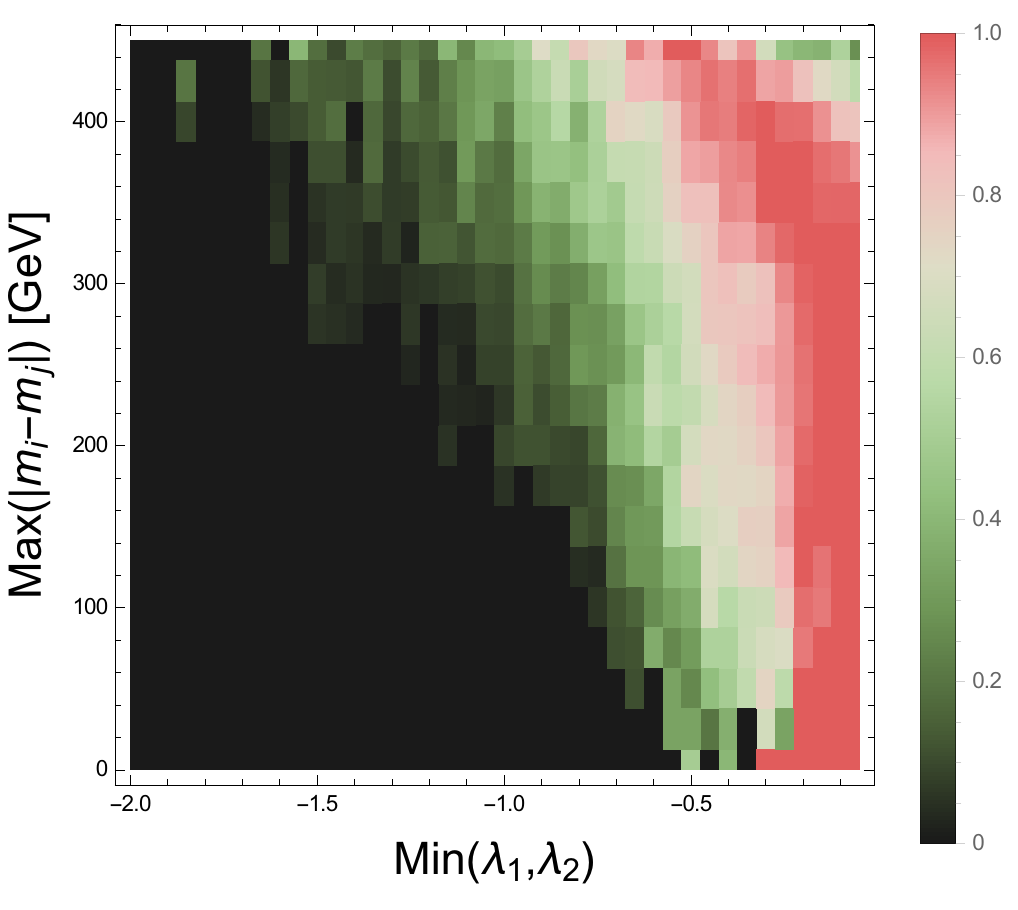} 
\caption{The misidentification rate $r$ as function of the maximal mass splitting of the heavy Higgs states and the size of $\lambda_{1,2}$.}
\label{fig:changeUFB}
\end{figure}

\begin{figure}[tb]
\includegraphics[width=\linewidth]{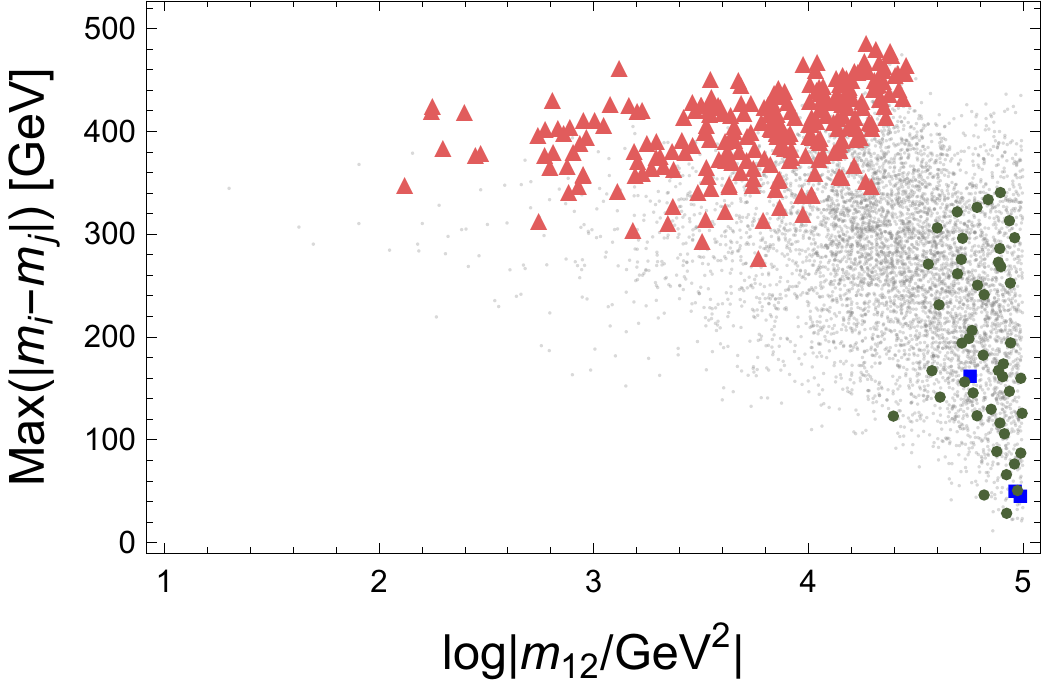} 
\caption{Distribution of points without UFB: they gray points (blue squares) are stable (unstable) at tree- and loop-level. The red triangles are stable at tree- but unstable at loop-level, while green points are unstable at tree- and stable at loop-level.}
\label{fig:changeMeta}
\end{figure}

As we have seen, loop corrections can be very important to judge the stability of the THDM. Therefore, we are going to check now how often this can happen in a common parameter scan. For that purpose, we use \Vevacious\cite{Camargo-Molina:2013qva} to test the stability of the one-loop effective potential. We have generated the necessary model files with \SARAH \footnote{We were using two model files. One with the possibility of additional charge and CP breaking VEVs, but found no difference compared to the results with only $v_1$, $v_2$.}. We also used \SARAH to generate a \SPheno module \cite{Porod:2003um,Porod:2011nf} for the THDM which use the masses and $\tan\alpha$ as input. \SPheno automatically translate this input into the tree-level couplings. In addition, we have modified the code to calculate also the CTs for the $\lambda$'s which are necessary to keep the loop masses to their tree-level values. This information is then passed to \Vevacious to check the vacuum stability of the one-loop effective potential. As data sample we have generated 400,000 points using the following parameter ranges\footnote{We use small values of $\tan\beta$ to improve the efficiency of the random scan. For larger values, it is more likely the the quartic couplings violate the tree-level unitarity limits unless fine-tuned cancellations between different mass terms are present, see eqs.~(\ref{eq:l1})--(\ref{eq:l5}). The overall results are not affected by this choice.}:
\begin{eqnarray*}
& 200~\text{GeV} < m_H, m_A < 1000~\text{GeV} & \\
&  500~\text{GeV} < m_{H^+} < 1000~\text{GeV} & \\
& -10^6~\text{GeV}^2 < m_{12} < 0 & \\
& -1 < \tan\alpha < 0,\quad 1 < \tan\beta < 1.5 &
\end{eqnarray*}
Afterwards, points are discarded which violate the tree-level unitarity limits or which fail the {\tt HiggsBounds} checks \cite{Bechtle:2008jh,Bechtle:2013wla}. 
The remaining 22,395 points can be categorised as shown in Tab.~\ref{tab:results}.
\begin{table}[tb]
\begin{tabular}{c|c|cc|c}
\hline
\hline
       &Tree  & \multicolumn{2}{|c|}{Loop} & r \\
       &      & Stable & Unstable &    \\
\hline       
UFB    & 15,975 & 9,157   & 6,818    & 57.3\% \\
Meta  & 51   & 48     & 3    & 94.1 \% \\
Stable & 6,369 & 6,116   & 253    & 4.0  \% \\
\hline
\hline
\end{tabular}
\caption{Summary of our parameter set. $r$ is the misidentification rate when using tree-level constraints. 'Unstable' at loop level includes UFB and
metastability, i.e. 'stable' means absolutely stable.}
\label{tab:results}
\end{table}
Thus, more than half of the points which are ruled out by the tree-level UFB checks are valid at the loop-level. In general, there is a correlation between the size of the quartic couplings and the mass splitting between $m_H$, $m_A$ and $m_{H^+}$. Consequently, we find also a correlation between the maximal splitting between the heavy Higgs state and the size of $\lambda_{1,2}$ which can be stabilised via loop corrections\footnote{This implies, of course, that the rate of misidentified points depends on the chosen parameter ranges and gets enhanced by the different lower limit of $m_{H^+}$ compared to $m_H$, $m_A$. Nevertheless, we think that the results are representative, because in literature often even bigger differences between the charged and heavy neutral masses are considered.}. This is shown in Fig.~\ref{fig:changeUFB} where the misidentification rate $r$ is given as function of 
$\text{min}(\lambda_1,\lambda_2)$ and the maximal mass difference. $r$ gives for the UFB and metastability check the ratio of points for which the result 'unstable' changes to 'stable' at the loop-level, while for stable tree-level points it is vice versa. For small ($>-0.2$) negative values of $\lambda_{1,2}$, we find $r\approx 1$  for the entire range of mass differences. Only for a small island with very large mass differences points stay unstable at the loop-level. These points have in common that $\lambda_3$ is  $O(4 \pi)$, i.e. the loop corrections might not be under control any more. If we would have applied a stronger cut on $|\lambda_i|$ as it might be necessary to really keep perturbativity under control \cite{Nierste:1995zx}, e.g. $|\lambda_i|<2\pi$, this island wouldn't appear and $r=1$ would hold up to $\text{min}(\lambda_1,\lambda_2)\gtrsim-0.15$. 
To test the usefulness of the check for metastability, the sample of points is significantly lower than for UFB, i.e. there is a non-negligible uncertainty in the misidentification rate. Nevertheless, the obtained results suggest that in most cases it rules out points which are viable. On the other side, the fraction of points which is stable at tree-level but becomes unstable at loop-level is quite low. We show the distribution of points without UFB directions in Fig.~\ref{fig:changeMeta}. One can see that only for small $|m_{12}|$ and large mass differences between the heavy Higgs states, a point stable at tree-level can become unstable at loop-level. \\

Up to now, we have only studied the overall stability of the potential. However, even a meta-stable vacuum is viable as long as the life-time exceeds the age of the universe. We have checked the points with deeper minima using the code {\tt CosmoTransition} \cite{Wainwright:2011kj}  and found that the majority of points has a comparable short life-time. Only in about 5\% of the cases, the tunnelling rate is sufficiently small to consider these points long-lived at zero temperature. If thermal corrections are included, the fraction of long-lived points shrinks to 1\%.  \\

\section{Conclusion}
\label{sec:conclusion}
We have studied in this letter the effects of radiative corrections to the vacuum stability conditions in THDMs. In these models large quartic couplings appear if large mass differences between the heavy Higgs states are considered. These large couplings cause important loop correction to the scalar potential. As consequence, we found that a large fraction of points which is ruled out by tree-level conditions are revived at the loop level. This happened in more than 50\% of the cases for points failing the standard UFB checks, and even in more than 90\% of the cases for the tree-level metastability check. Because of the importance of the UFB checks, more than 40\% of all phenomenological viable points are misidentified at tree-level. If no checks for vacuum stability would have been applied at all, the fraction of wrong points would be only $\sim 30\%$ for the considered dataset.
Because of these large misidentification rates, it seems necessary to push the standards of these theoretical constraints beyond the tree-level. It is also very likely that similar results would be found for other non-supersymmetric models if quartic couplings $\gg 1$ are used. \\


\section*{Acknowledgements}
I'm grateful to  David Lopez-Val who patiently answered my basic questions about THDMs.
I thank Johannes Braathen, Mark D. Goodsell, Manuel E. Krauss, Moritz Laber, Maggie M\"uhlleitner and Toby Opferkuch for various discussions which triggered my interest in the vacuum stability checks in non-supersymmetric models. 

\bibliography{lit}

\begin{thebibliography}{62}%
\makeatletter
\providecommand \@ifxundefined [1]{%
 \@ifx{#1\undefined}
}%
\providecommand \@ifnum [1]{%
 \ifnum #1\expandafter \@firstoftwo
 \else \expandafter \@secondoftwo
 \fi
}%
\providecommand \@ifx [1]{%
 \ifx #1\expandafter \@firstoftwo
 \else \expandafter \@secondoftwo
 \fi
}%
\providecommand \natexlab [1]{#1}%
\providecommand \enquote  [1]{``#1''}%
\providecommand \bibnamefont  [1]{#1}%
\providecommand \bibfnamefont [1]{#1}%
\providecommand \citenamefont [1]{#1}%
\providecommand \href@noop [0]{\@secondoftwo}%
\providecommand \href [0]{\begingroup \@sanitize@url \@href}%
\providecommand \@href[1]{\@@startlink{#1}\@@href}%
\providecommand \@@href[1]{\endgroup#1\@@endlink}%
\providecommand \@sanitize@url [0]{\catcode `\\12\catcode `\$12\catcode
  `\&12\catcode `\#12\catcode `\^12\catcode `\_12\catcode `\%12\relax}%
\providecommand \@@startlink[1]{}%
\providecommand \@@endlink[0]{}%
\providecommand \url  [0]{\begingroup\@sanitize@url \@url }%
\providecommand \@url [1]{\endgroup\@href {#1}{\urlprefix }}%
\providecommand \urlprefix  [0]{URL }%
\providecommand \Eprint [0]{\href }%
\providecommand \doibase [0]{http://dx.doi.org/}%
\providecommand \selectlanguage [0]{\@gobble}%
\providecommand \bibinfo  [0]{\@secondoftwo}%
\providecommand \bibfield  [0]{\@secondoftwo}%
\providecommand \translation [1]{[#1]}%
\providecommand \BibitemOpen [0]{}%
\providecommand \bibitemStop [0]{}%
\providecommand \bibitemNoStop [0]{.\EOS\space}%
\providecommand \EOS [0]{\spacefactor3000\relax}%
\providecommand \BibitemShut  [1]{\csname bibitem#1\endcsname}%
\let\auto@bib@innerbib\@empty
\bibitem [{\citenamefont {Aad}\ \emph {et~al.}(2012)\citenamefont {Aad} \emph
  {et~al.}}]{Aad:2012tfa}%
  \BibitemOpen
  \bibfield  {author} {\bibinfo {author} {\bibfnamefont {G.}~\bibnamefont
  {Aad}} \emph {et~al.} (\bibinfo {collaboration} {ATLAS Collaboration}),\
  }\href {\doibase 10.1016/j.physletb.2012.08.020} {\bibfield  {journal}
  {\bibinfo  {journal} {Phys.Lett.}\ }\textbf {\bibinfo {volume} {B716}},\
  \bibinfo {pages} {1} (\bibinfo {year} {2012})},\ \Eprint
  {http://arxiv.org/abs/1207.7214} {arXiv:1207.7214 [hep-ex]} \BibitemShut
  {NoStop}%
\bibitem [{\citenamefont {Chatrchyan}\ \emph {et~al.}(2012)\citenamefont
  {Chatrchyan} \emph {et~al.}}]{Chatrchyan:2012xdj}%
  \BibitemOpen
  \bibfield  {author} {\bibinfo {author} {\bibfnamefont {S.}~\bibnamefont
  {Chatrchyan}} \emph {et~al.} (\bibinfo {collaboration} {CMS}),\ }\href
  {\doibase 10.1016/j.physletb.2012.08.021} {\bibfield  {journal} {\bibinfo
  {journal} {Phys. Lett.}\ }\textbf {\bibinfo {volume} {B716}},\ \bibinfo
  {pages} {30} (\bibinfo {year} {2012})},\ \Eprint
  {http://arxiv.org/abs/1207.7235} {arXiv:1207.7235 [hep-ex]} \BibitemShut
  {NoStop}%
\bibitem [{\citenamefont {Degrassi}\ \emph {et~al.}(2012)\citenamefont
  {Degrassi}, \citenamefont {Di~Vita}, \citenamefont {Elias-Miro},
  \citenamefont {Espinosa}, \citenamefont {Giudice}, \citenamefont {Isidori},\
  and\ \citenamefont {Strumia}}]{Degrassi:2012ry}%
  \BibitemOpen
  \bibfield  {author} {\bibinfo {author} {\bibfnamefont {G.}~\bibnamefont
  {Degrassi}}, \bibinfo {author} {\bibfnamefont {S.}~\bibnamefont {Di~Vita}},
  \bibinfo {author} {\bibfnamefont {J.}~\bibnamefont {Elias-Miro}}, \bibinfo
  {author} {\bibfnamefont {J.~R.}\ \bibnamefont {Espinosa}}, \bibinfo {author}
  {\bibfnamefont {G.~F.}\ \bibnamefont {Giudice}}, \bibinfo {author}
  {\bibfnamefont {G.}~\bibnamefont {Isidori}}, \ and\ \bibinfo {author}
  {\bibfnamefont {A.}~\bibnamefont {Strumia}},\ }\href {\doibase
  10.1007/JHEP08(2012)098} {\bibfield  {journal} {\bibinfo  {journal} {JHEP}\
  }\textbf {\bibinfo {volume} {08}},\ \bibinfo {pages} {098} (\bibinfo {year}
  {2012})},\ \Eprint {http://arxiv.org/abs/1205.6497} {arXiv:1205.6497
  [hep-ph]} \BibitemShut {NoStop}%
\bibitem [{\citenamefont {Nilles}\ \emph {et~al.}(1983)\citenamefont {Nilles},
  \citenamefont {Srednicki},\ and\ \citenamefont {Wyler}}]{Nilles:1982dy}%
  \BibitemOpen
  \bibfield  {author} {\bibinfo {author} {\bibfnamefont {H.~P.}\ \bibnamefont
  {Nilles}}, \bibinfo {author} {\bibfnamefont {M.}~\bibnamefont {Srednicki}}, \
  and\ \bibinfo {author} {\bibfnamefont {D.}~\bibnamefont {Wyler}},\ }\href
  {\doibase 10.1016/0370-2693(83)90460-4} {\bibfield  {journal} {\bibinfo
  {journal} {Phys. Lett.}\ }\textbf {\bibinfo {volume} {B120}},\ \bibinfo
  {pages} {346} (\bibinfo {year} {1983})}\BibitemShut {NoStop}%
\bibitem [{\citenamefont {Alvarez-Gaume}\ \emph {et~al.}(1983)\citenamefont
  {Alvarez-Gaume}, \citenamefont {Polchinski},\ and\ \citenamefont
  {Wise}}]{AlvarezGaume:1983gj}%
  \BibitemOpen
  \bibfield  {author} {\bibinfo {author} {\bibfnamefont {L.}~\bibnamefont
  {Alvarez-Gaume}}, \bibinfo {author} {\bibfnamefont {J.}~\bibnamefont
  {Polchinski}}, \ and\ \bibinfo {author} {\bibfnamefont {M.~B.}\ \bibnamefont
  {Wise}},\ }\href {\doibase 10.1016/0550-3213(83)90591-6} {\bibfield
  {journal} {\bibinfo  {journal} {Nucl. Phys.}\ }\textbf {\bibinfo {volume}
  {B221}},\ \bibinfo {pages} {495} (\bibinfo {year} {1983})}\BibitemShut
  {NoStop}%
\bibitem [{\citenamefont {Derendinger}\ and\ \citenamefont
  {Savoy}(1984)}]{Derendinger:1983bz}%
  \BibitemOpen
  \bibfield  {author} {\bibinfo {author} {\bibfnamefont {J.~P.}\ \bibnamefont
  {Derendinger}}\ and\ \bibinfo {author} {\bibfnamefont {C.~A.}\ \bibnamefont
  {Savoy}},\ }\href {\doibase 10.1016/0550-3213(84)90162-7} {\bibfield
  {journal} {\bibinfo  {journal} {Nucl. Phys.}\ }\textbf {\bibinfo {volume}
  {B237}},\ \bibinfo {pages} {307} (\bibinfo {year} {1984})}\BibitemShut
  {NoStop}%
\bibitem [{\citenamefont {Claudson}\ \emph {et~al.}(1983)\citenamefont
  {Claudson}, \citenamefont {Hall},\ and\ \citenamefont
  {Hinchliffe}}]{Claudson:1983et}%
  \BibitemOpen
  \bibfield  {author} {\bibinfo {author} {\bibfnamefont {M.}~\bibnamefont
  {Claudson}}, \bibinfo {author} {\bibfnamefont {L.~J.}\ \bibnamefont {Hall}},
  \ and\ \bibinfo {author} {\bibfnamefont {I.}~\bibnamefont {Hinchliffe}},\
  }\href {\doibase 10.1016/0550-3213(83)90556-4} {\bibfield  {journal}
  {\bibinfo  {journal} {Nucl. Phys.}\ }\textbf {\bibinfo {volume} {B228}},\
  \bibinfo {pages} {501} (\bibinfo {year} {1983})}\BibitemShut {NoStop}%
\bibitem [{\citenamefont {Kounnas}\ \emph {et~al.}(1984)\citenamefont
  {Kounnas}, \citenamefont {Lahanas}, \citenamefont {Nanopoulos},\ and\
  \citenamefont {Quiros}}]{Kounnas:1983td}%
  \BibitemOpen
  \bibfield  {author} {\bibinfo {author} {\bibfnamefont {C.}~\bibnamefont
  {Kounnas}}, \bibinfo {author} {\bibfnamefont {A.~B.}\ \bibnamefont
  {Lahanas}}, \bibinfo {author} {\bibfnamefont {D.~V.}\ \bibnamefont
  {Nanopoulos}}, \ and\ \bibinfo {author} {\bibfnamefont {M.}~\bibnamefont
  {Quiros}},\ }\href {\doibase 10.1016/0550-3213(84)90545-5} {\bibfield
  {journal} {\bibinfo  {journal} {Nucl. Phys.}\ }\textbf {\bibinfo {volume}
  {B236}},\ \bibinfo {pages} {438} (\bibinfo {year} {1984})}\BibitemShut
  {NoStop}%
\bibitem [{\citenamefont {Drees}\ \emph {et~al.}(1985)\citenamefont {Drees},
  \citenamefont {Gluck},\ and\ \citenamefont {Grassie}}]{Drees:1985ie}%
  \BibitemOpen
  \bibfield  {author} {\bibinfo {author} {\bibfnamefont {M.}~\bibnamefont
  {Drees}}, \bibinfo {author} {\bibfnamefont {M.}~\bibnamefont {Gluck}}, \ and\
  \bibinfo {author} {\bibfnamefont {K.}~\bibnamefont {Grassie}},\ }\href
  {\doibase 10.1016/0370-2693(85)91538-2} {\bibfield  {journal} {\bibinfo
  {journal} {Phys. Lett.}\ }\textbf {\bibinfo {volume} {B157}},\ \bibinfo
  {pages} {164} (\bibinfo {year} {1985})}\BibitemShut {NoStop}%
\bibitem [{\citenamefont {Gunion}\ \emph {et~al.}(1988)\citenamefont {Gunion},
  \citenamefont {Haber},\ and\ \citenamefont {Sher}}]{Gunion:1987qv}%
  \BibitemOpen
  \bibfield  {author} {\bibinfo {author} {\bibfnamefont {J.~F.}\ \bibnamefont
  {Gunion}}, \bibinfo {author} {\bibfnamefont {H.~E.}\ \bibnamefont {Haber}}, \
  and\ \bibinfo {author} {\bibfnamefont {M.}~\bibnamefont {Sher}},\ }\href
  {\doibase 10.1016/0550-3213(88)90168-X} {\bibfield  {journal} {\bibinfo
  {journal} {Nucl. Phys.}\ }\textbf {\bibinfo {volume} {B306}},\ \bibinfo
  {pages} {1} (\bibinfo {year} {1988})}\BibitemShut {NoStop}%
\bibitem [{\citenamefont {Komatsu}(1988)}]{Komatsu:1988mt}%
  \BibitemOpen
  \bibfield  {author} {\bibinfo {author} {\bibfnamefont {H.}~\bibnamefont
  {Komatsu}},\ }\href {\doibase 10.1016/0370-2693(88)91441-4} {\bibfield
  {journal} {\bibinfo  {journal} {Phys. Lett.}\ }\textbf {\bibinfo {volume}
  {B215}},\ \bibinfo {pages} {323} (\bibinfo {year} {1988})}\BibitemShut
  {NoStop}%
\bibitem [{\citenamefont {Langacker}\ and\ \citenamefont
  {Polonsky}(1994)}]{Langacker:1994bc}%
  \BibitemOpen
  \bibfield  {author} {\bibinfo {author} {\bibfnamefont {P.}~\bibnamefont
  {Langacker}}\ and\ \bibinfo {author} {\bibfnamefont {N.}~\bibnamefont
  {Polonsky}},\ }\href {\doibase 10.1103/PhysRevD.50.2199} {\bibfield
  {journal} {\bibinfo  {journal} {Phys. Rev.}\ }\textbf {\bibinfo {volume}
  {D50}},\ \bibinfo {pages} {2199} (\bibinfo {year} {1994})},\ \Eprint
  {http://arxiv.org/abs/hep-ph/9403306} {arXiv:hep-ph/9403306 [hep-ph]}
  \BibitemShut {NoStop}%
\bibitem [{\citenamefont {Casas}\ \emph {et~al.}(1996)\citenamefont {Casas},
  \citenamefont {Lleyda},\ and\ \citenamefont {Munoz}}]{Casas:1995pd}%
  \BibitemOpen
  \bibfield  {author} {\bibinfo {author} {\bibfnamefont {J.~A.}\ \bibnamefont
  {Casas}}, \bibinfo {author} {\bibfnamefont {A.}~\bibnamefont {Lleyda}}, \
  and\ \bibinfo {author} {\bibfnamefont {C.}~\bibnamefont {Munoz}},\ }\href
  {\doibase 10.1016/0550-3213(96)00194-0} {\bibfield  {journal} {\bibinfo
  {journal} {Nucl. Phys.}\ }\textbf {\bibinfo {volume} {B471}},\ \bibinfo
  {pages} {3} (\bibinfo {year} {1996})},\ \Eprint
  {http://arxiv.org/abs/hep-ph/9507294} {arXiv:hep-ph/9507294 [hep-ph]}
  \BibitemShut {NoStop}%
\bibitem [{\citenamefont {Casas}\ and\ \citenamefont
  {Dimopoulos}(1996)}]{Casas:1996de}%
  \BibitemOpen
  \bibfield  {author} {\bibinfo {author} {\bibfnamefont {J.~A.}\ \bibnamefont
  {Casas}}\ and\ \bibinfo {author} {\bibfnamefont {S.}~\bibnamefont
  {Dimopoulos}},\ }\href {\doibase 10.1016/0370-2693(96)01000-3} {\bibfield
  {journal} {\bibinfo  {journal} {Phys. Lett.}\ }\textbf {\bibinfo {volume}
  {B387}},\ \bibinfo {pages} {107} (\bibinfo {year} {1996})},\ \Eprint
  {http://arxiv.org/abs/hep-ph/9606237} {arXiv:hep-ph/9606237 [hep-ph]}
  \BibitemShut {NoStop}%
\bibitem [{\citenamefont {Camargo-Molina}\ \emph
  {et~al.}(2013{\natexlab{a}})\citenamefont {Camargo-Molina}, \citenamefont
  {O'Leary}, \citenamefont {Porod},\ and\ \citenamefont
  {Staub}}]{CamargoMolina:2012hv}%
  \BibitemOpen
  \bibfield  {author} {\bibinfo {author} {\bibfnamefont {J.~E.}\ \bibnamefont
  {Camargo-Molina}}, \bibinfo {author} {\bibfnamefont {B.}~\bibnamefont
  {O'Leary}}, \bibinfo {author} {\bibfnamefont {W.}~\bibnamefont {Porod}}, \
  and\ \bibinfo {author} {\bibfnamefont {F.}~\bibnamefont {Staub}},\ }\href
  {\doibase 10.1103/PhysRevD.88.015033} {\bibfield  {journal} {\bibinfo
  {journal} {Phys. Rev.}\ }\textbf {\bibinfo {volume} {D88}},\ \bibinfo {pages}
  {015033} (\bibinfo {year} {2013}{\natexlab{a}})},\ \Eprint
  {http://arxiv.org/abs/1212.4146} {arXiv:1212.4146 [hep-ph]} \BibitemShut
  {NoStop}%
\bibitem [{\citenamefont {Camargo-Molina}\ \emph
  {et~al.}(2013{\natexlab{b}})\citenamefont {Camargo-Molina}, \citenamefont
  {O'Leary}, \citenamefont {Porod},\ and\ \citenamefont
  {Staub}}]{Camargo-Molina:2013sta}%
  \BibitemOpen
  \bibfield  {author} {\bibinfo {author} {\bibfnamefont {J.~E.}\ \bibnamefont
  {Camargo-Molina}}, \bibinfo {author} {\bibfnamefont {B.}~\bibnamefont
  {O'Leary}}, \bibinfo {author} {\bibfnamefont {W.}~\bibnamefont {Porod}}, \
  and\ \bibinfo {author} {\bibfnamefont {F.}~\bibnamefont {Staub}},\ }\href
  {\doibase 10.1007/JHEP12(2013)103} {\bibfield  {journal} {\bibinfo  {journal}
  {JHEP}\ }\textbf {\bibinfo {volume} {12}},\ \bibinfo {pages} {103} (\bibinfo
  {year} {2013}{\natexlab{b}})},\ \Eprint {http://arxiv.org/abs/1309.7212}
  {arXiv:1309.7212 [hep-ph]} \BibitemShut {NoStop}%
\bibitem [{\citenamefont {Chowdhury}\ \emph {et~al.}(2014)\citenamefont
  {Chowdhury}, \citenamefont {Godbole}, \citenamefont {Mohan},\ and\
  \citenamefont {Vempati}}]{Chowdhury:2013dka}%
  \BibitemOpen
  \bibfield  {author} {\bibinfo {author} {\bibfnamefont {D.}~\bibnamefont
  {Chowdhury}}, \bibinfo {author} {\bibfnamefont {R.~M.}\ \bibnamefont
  {Godbole}}, \bibinfo {author} {\bibfnamefont {K.~A.}\ \bibnamefont {Mohan}},
  \ and\ \bibinfo {author} {\bibfnamefont {S.~K.}\ \bibnamefont {Vempati}},\
  }\href {\doibase 10.1007/JHEP02(2014)110} {\bibfield  {journal} {\bibinfo
  {journal} {JHEP}\ }\textbf {\bibinfo {volume} {02}},\ \bibinfo {pages} {110}
  (\bibinfo {year} {2014})},\ \Eprint {http://arxiv.org/abs/1310.1932}
  {arXiv:1310.1932 [hep-ph]} \BibitemShut {NoStop}%
\bibitem [{\citenamefont {Blinov}\ and\ \citenamefont
  {Morrissey}(2014)}]{Blinov:2013fta}%
  \BibitemOpen
  \bibfield  {author} {\bibinfo {author} {\bibfnamefont {N.}~\bibnamefont
  {Blinov}}\ and\ \bibinfo {author} {\bibfnamefont {D.~E.}\ \bibnamefont
  {Morrissey}},\ }\href {\doibase 10.1007/JHEP03(2014)106} {\bibfield
  {journal} {\bibinfo  {journal} {JHEP}\ }\textbf {\bibinfo {volume} {03}},\
  \bibinfo {pages} {106} (\bibinfo {year} {2014})},\ \Eprint
  {http://arxiv.org/abs/1310.4174} {arXiv:1310.4174 [hep-ph]} \BibitemShut
  {NoStop}%
\bibitem [{\citenamefont {Camargo-Molina}\ \emph
  {et~al.}(2013{\natexlab{c}})\citenamefont {Camargo-Molina}, \citenamefont
  {O'Leary}, \citenamefont {Porod},\ and\ \citenamefont
  {Staub}}]{MOLINA:2014uha}%
  \BibitemOpen
  \bibfield  {author} {\bibinfo {author} {\bibfnamefont {J.~E.}\ \bibnamefont
  {Camargo-Molina}}, \bibinfo {author} {\bibfnamefont {B.}~\bibnamefont
  {O'Leary}}, \bibinfo {author} {\bibfnamefont {W.}~\bibnamefont {Porod}}, \
  and\ \bibinfo {author} {\bibfnamefont {F.}~\bibnamefont {Staub}},\ }\bibfield
   {booktitle} {\emph {\bibinfo {booktitle} {{EPS-HEP 2013}}},\ }\href@noop {}
  {\bibfield  {journal} {\bibinfo  {journal} {PoS}\ }\textbf {\bibinfo {volume}
  {EPS-HEP2013}},\ \bibinfo {pages} {265} (\bibinfo {year}
  {2013}{\natexlab{c}})},\ \Eprint {http://arxiv.org/abs/1310.1260}
  {arXiv:1310.1260 [hep-ph]} \BibitemShut {NoStop}%
\bibitem [{\citenamefont {Chattopadhyay}\ and\ \citenamefont
  {Dey}(2014)}]{Chattopadhyay:2014gfa}%
  \BibitemOpen
  \bibfield  {author} {\bibinfo {author} {\bibfnamefont {U.}~\bibnamefont
  {Chattopadhyay}}\ and\ \bibinfo {author} {\bibfnamefont {A.}~\bibnamefont
  {Dey}},\ }\href {\doibase 10.1007/JHEP11(2014)161} {\bibfield  {journal}
  {\bibinfo  {journal} {JHEP}\ }\textbf {\bibinfo {volume} {11}},\ \bibinfo
  {pages} {161} (\bibinfo {year} {2014})},\ \Eprint
  {http://arxiv.org/abs/1409.0611} {arXiv:1409.0611 [hep-ph]} \BibitemShut
  {NoStop}%
\bibitem [{\citenamefont {Camargo-Molina}\ \emph {et~al.}(2014)\citenamefont
  {Camargo-Molina}, \citenamefont {Garbrecht}, \citenamefont {O'Leary},
  \citenamefont {Porod},\ and\ \citenamefont {Staub}}]{Camargo-Molina:2014pwa}%
  \BibitemOpen
  \bibfield  {author} {\bibinfo {author} {\bibfnamefont {J.~E.}\ \bibnamefont
  {Camargo-Molina}}, \bibinfo {author} {\bibfnamefont {B.}~\bibnamefont
  {Garbrecht}}, \bibinfo {author} {\bibfnamefont {B.}~\bibnamefont {O'Leary}},
  \bibinfo {author} {\bibfnamefont {W.}~\bibnamefont {Porod}}, \ and\ \bibinfo
  {author} {\bibfnamefont {F.}~\bibnamefont {Staub}},\ }\href {\doibase
  10.1016/j.physletb.2014.08.036} {\bibfield  {journal} {\bibinfo  {journal}
  {Phys. Lett.}\ }\textbf {\bibinfo {volume} {B737}},\ \bibinfo {pages} {156}
  (\bibinfo {year} {2014})},\ \Eprint {http://arxiv.org/abs/1405.7376}
  {arXiv:1405.7376 [hep-ph]} \BibitemShut {NoStop}%
\bibitem [{\citenamefont {Hollik}(2016)}]{Hollik:2016dcm}%
  \BibitemOpen
  \bibfield  {author} {\bibinfo {author} {\bibfnamefont {W.~G.}\ \bibnamefont
  {Hollik}},\ }\href {\doibase 10.1007/JHEP08(2016)126} {\bibfield  {journal}
  {\bibinfo  {journal} {JHEP}\ }\textbf {\bibinfo {volume} {08}},\ \bibinfo
  {pages} {126} (\bibinfo {year} {2016})},\ \Eprint
  {http://arxiv.org/abs/1606.08356} {arXiv:1606.08356 [hep-ph]} \BibitemShut
  {NoStop}%
\bibitem [{\citenamefont {Beuria}\ \emph {et~al.}(2016)\citenamefont {Beuria},
  \citenamefont {Chattopadhyay}, \citenamefont {Datta},\ and\ \citenamefont
  {Dey}}]{Beuria:2016cdk}%
  \BibitemOpen
  \bibfield  {author} {\bibinfo {author} {\bibfnamefont {J.}~\bibnamefont
  {Beuria}}, \bibinfo {author} {\bibfnamefont {U.}~\bibnamefont
  {Chattopadhyay}}, \bibinfo {author} {\bibfnamefont {A.}~\bibnamefont
  {Datta}}, \ and\ \bibinfo {author} {\bibfnamefont {A.}~\bibnamefont {Dey}},\
  }\href@noop {} {\  (\bibinfo {year} {2016})},\ \Eprint
  {http://arxiv.org/abs/1612.06803} {arXiv:1612.06803 [hep-ph]} \BibitemShut
  {NoStop}%
\bibitem [{\citenamefont {Dreiner}\ \emph {et~al.}(2016)\citenamefont
  {Dreiner}, \citenamefont {Krauss}, \citenamefont {O'Leary}, \citenamefont
  {Opferkuch},\ and\ \citenamefont {Staub}}]{Dreiner:2016wwk}%
  \BibitemOpen
  \bibfield  {author} {\bibinfo {author} {\bibfnamefont {H.~K.}\ \bibnamefont
  {Dreiner}}, \bibinfo {author} {\bibfnamefont {M.~E.}\ \bibnamefont {Krauss}},
  \bibinfo {author} {\bibfnamefont {B.}~\bibnamefont {O'Leary}}, \bibinfo
  {author} {\bibfnamefont {T.}~\bibnamefont {Opferkuch}}, \ and\ \bibinfo
  {author} {\bibfnamefont {F.}~\bibnamefont {Staub}},\ }\href {\doibase
  10.1103/PhysRevD.94.055013} {\bibfield  {journal} {\bibinfo  {journal} {Phys.
  Rev.}\ }\textbf {\bibinfo {volume} {D94}},\ \bibinfo {pages} {055013}
  (\bibinfo {year} {2016})},\ \Eprint {http://arxiv.org/abs/1606.08811}
  {arXiv:1606.08811 [hep-ph]} \BibitemShut {NoStop}%
\bibitem [{\citenamefont {Krauss}\ \emph {et~al.}(2017)\citenamefont {Krauss},
  \citenamefont {Opferkuch},\ and\ \citenamefont {Staub}}]{Krauss:2017nlh}%
  \BibitemOpen
  \bibfield  {author} {\bibinfo {author} {\bibfnamefont {M.~E.}\ \bibnamefont
  {Krauss}}, \bibinfo {author} {\bibfnamefont {T.}~\bibnamefont {Opferkuch}}, \
  and\ \bibinfo {author} {\bibfnamefont {F.}~\bibnamefont {Staub}},\
  }\href@noop {} {\  (\bibinfo {year} {2017})},\ \Eprint
  {http://arxiv.org/abs/1703.05329} {arXiv:1703.05329 [hep-ph]} \BibitemShut
  {NoStop}%
\bibitem [{\citenamefont {Klimenko}(1985)}]{Klimenko:1984qx}%
  \BibitemOpen
  \bibfield  {author} {\bibinfo {author} {\bibfnamefont {K.~G.}\ \bibnamefont
  {Klimenko}},\ }\href {\doibase 10.1007/BF01034825} {\bibfield  {journal}
  {\bibinfo  {journal} {Theor. Math. Phys.}\ }\textbf {\bibinfo {volume}
  {62}},\ \bibinfo {pages} {58} (\bibinfo {year} {1985})},\ \bibinfo {note}
  {[Teor. Mat. Fiz.62,87(1985)]}\BibitemShut {NoStop}%
\bibitem [{\citenamefont {Velhinho}\ \emph {et~al.}(1994)\citenamefont
  {Velhinho}, \citenamefont {Santos},\ and\ \citenamefont
  {Barroso}}]{Velhinho:1994np}%
  \BibitemOpen
  \bibfield  {author} {\bibinfo {author} {\bibfnamefont {J.}~\bibnamefont
  {Velhinho}}, \bibinfo {author} {\bibfnamefont {R.}~\bibnamefont {Santos}}, \
  and\ \bibinfo {author} {\bibfnamefont {A.}~\bibnamefont {Barroso}},\ }\href
  {\doibase 10.1016/0370-2693(94)91109-6} {\bibfield  {journal} {\bibinfo
  {journal} {Phys. Lett.}\ }\textbf {\bibinfo {volume} {B322}},\ \bibinfo
  {pages} {213} (\bibinfo {year} {1994})}\BibitemShut {NoStop}%
\bibitem [{\citenamefont {Ferreira}\ \emph {et~al.}(2004)\citenamefont
  {Ferreira}, \citenamefont {Santos},\ and\ \citenamefont
  {Barroso}}]{Ferreira:2004yd}%
  \BibitemOpen
  \bibfield  {author} {\bibinfo {author} {\bibfnamefont {P.~M.}\ \bibnamefont
  {Ferreira}}, \bibinfo {author} {\bibfnamefont {R.}~\bibnamefont {Santos}}, \
  and\ \bibinfo {author} {\bibfnamefont {A.}~\bibnamefont {Barroso}},\ }\href
  {\doibase 10.1016/j.physletb.2004.10.022, 10.1016/j.physletb.2005.09.074}
  {\bibfield  {journal} {\bibinfo  {journal} {Phys. Lett.}\ }\textbf {\bibinfo
  {volume} {B603}},\ \bibinfo {pages} {219} (\bibinfo {year} {2004})},\
  \bibinfo {note} {[Erratum: Phys. Lett.B629,114(2005)]},\ \Eprint
  {http://arxiv.org/abs/hep-ph/0406231} {arXiv:hep-ph/0406231 [hep-ph]}
  \BibitemShut {NoStop}%
\bibitem [{\citenamefont {Barroso}\ \emph {et~al.}(2006)\citenamefont
  {Barroso}, \citenamefont {Ferreira},\ and\ \citenamefont
  {Santos}}]{Barroso:2005sm}%
  \BibitemOpen
  \bibfield  {author} {\bibinfo {author} {\bibfnamefont {A.}~\bibnamefont
  {Barroso}}, \bibinfo {author} {\bibfnamefont {P.~M.}\ \bibnamefont
  {Ferreira}}, \ and\ \bibinfo {author} {\bibfnamefont {R.}~\bibnamefont
  {Santos}},\ }\href {\doibase 10.1016/j.physletb.2005.11.031} {\bibfield
  {journal} {\bibinfo  {journal} {Phys. Lett.}\ }\textbf {\bibinfo {volume}
  {B632}},\ \bibinfo {pages} {684} (\bibinfo {year} {2006})},\ \Eprint
  {http://arxiv.org/abs/hep-ph/0507224} {arXiv:hep-ph/0507224 [hep-ph]}
  \BibitemShut {NoStop}%
\bibitem [{\citenamefont {Maniatis}\ \emph {et~al.}(2006)\citenamefont
  {Maniatis}, \citenamefont {von Manteuffel}, \citenamefont {Nachtmann},\ and\
  \citenamefont {Nagel}}]{Maniatis:2006fs}%
  \BibitemOpen
  \bibfield  {author} {\bibinfo {author} {\bibfnamefont {M.}~\bibnamefont
  {Maniatis}}, \bibinfo {author} {\bibfnamefont {A.}~\bibnamefont {von
  Manteuffel}}, \bibinfo {author} {\bibfnamefont {O.}~\bibnamefont
  {Nachtmann}}, \ and\ \bibinfo {author} {\bibfnamefont {F.}~\bibnamefont
  {Nagel}},\ }\href {\doibase 10.1140/epjc/s10052-006-0016-6} {\bibfield
  {journal} {\bibinfo  {journal} {Eur. Phys. J.}\ }\textbf {\bibinfo {volume}
  {C48}},\ \bibinfo {pages} {805} (\bibinfo {year} {2006})},\ \Eprint
  {http://arxiv.org/abs/hep-ph/0605184} {arXiv:hep-ph/0605184 [hep-ph]}
  \BibitemShut {NoStop}%
\bibitem [{\citenamefont {Ivanov}(2007)}]{Ivanov:2006yq}%
  \BibitemOpen
  \bibfield  {author} {\bibinfo {author} {\bibfnamefont {I.~P.}\ \bibnamefont
  {Ivanov}},\ }\href {\doibase 10.1103/PhysRevD.76.039902,
  10.1103/PhysRevD.75.035001} {\bibfield  {journal} {\bibinfo  {journal} {Phys.
  Rev.}\ }\textbf {\bibinfo {volume} {D75}},\ \bibinfo {pages} {035001}
  (\bibinfo {year} {2007})},\ \bibinfo {note} {[Erratum: Phys.
  Rev.D76,039902(2007)]},\ \Eprint {http://arxiv.org/abs/hep-ph/0609018}
  {arXiv:hep-ph/0609018 [hep-ph]} \BibitemShut {NoStop}%
\bibitem [{\citenamefont {Ivanov}(2008)}]{Ivanov:2007de}%
  \BibitemOpen
  \bibfield  {author} {\bibinfo {author} {\bibfnamefont {I.~P.}\ \bibnamefont
  {Ivanov}},\ }\href {\doibase 10.1103/PhysRevD.77.015017} {\bibfield
  {journal} {\bibinfo  {journal} {Phys. Rev.}\ }\textbf {\bibinfo {volume}
  {D77}},\ \bibinfo {pages} {015017} (\bibinfo {year} {2008})},\ \Eprint
  {http://arxiv.org/abs/0710.3490} {arXiv:0710.3490 [hep-ph]} \BibitemShut
  {NoStop}%
\bibitem [{\citenamefont {Ivanov}(2009)}]{Ivanov:2008er}%
  \BibitemOpen
  \bibfield  {author} {\bibinfo {author} {\bibfnamefont {I.~P.}\ \bibnamefont
  {Ivanov}},\ }\href@noop {} {\bibfield  {journal} {\bibinfo  {journal} {Acta
  Phys. Polon.}\ }\textbf {\bibinfo {volume} {B40}},\ \bibinfo {pages} {2789}
  (\bibinfo {year} {2009})},\ \Eprint {http://arxiv.org/abs/0812.4984}
  {arXiv:0812.4984 [hep-ph]} \BibitemShut {NoStop}%
\bibitem [{\citenamefont {Ginzburg}\ \emph {et~al.}(2010)\citenamefont
  {Ginzburg}, \citenamefont {Ivanov},\ and\ \citenamefont
  {Kanishev}}]{Ginzburg:2009dp}%
  \BibitemOpen
  \bibfield  {author} {\bibinfo {author} {\bibfnamefont {I.~F.}\ \bibnamefont
  {Ginzburg}}, \bibinfo {author} {\bibfnamefont {I.~P.}\ \bibnamefont
  {Ivanov}}, \ and\ \bibinfo {author} {\bibfnamefont {K.~A.}\ \bibnamefont
  {Kanishev}},\ }\href {\doibase 10.1103/PhysRevD.81.085031} {\bibfield
  {journal} {\bibinfo  {journal} {Phys. Rev.}\ }\textbf {\bibinfo {volume}
  {D81}},\ \bibinfo {pages} {085031} (\bibinfo {year} {2010})},\ \Eprint
  {http://arxiv.org/abs/0911.2383} {arXiv:0911.2383 [hep-ph]} \BibitemShut
  {NoStop}%
\bibitem [{\citenamefont {Ivanov}\ and\ \citenamefont
  {Nishi}(2010)}]{Ivanov:2010ww}%
  \BibitemOpen
  \bibfield  {author} {\bibinfo {author} {\bibfnamefont {I.~P.}\ \bibnamefont
  {Ivanov}}\ and\ \bibinfo {author} {\bibfnamefont {C.~C.}\ \bibnamefont
  {Nishi}},\ }\href {\doibase 10.1103/PhysRevD.82.015014} {\bibfield  {journal}
  {\bibinfo  {journal} {Phys. Rev.}\ }\textbf {\bibinfo {volume} {D82}},\
  \bibinfo {pages} {015014} (\bibinfo {year} {2010})},\ \Eprint
  {http://arxiv.org/abs/1004.1799} {arXiv:1004.1799 [hep-th]} \BibitemShut
  {NoStop}%
\bibitem [{\citenamefont {Ivanov}(2010)}]{Ivanov:2010wz}%
  \BibitemOpen
  \bibfield  {author} {\bibinfo {author} {\bibfnamefont {I.~P.}\ \bibnamefont
  {Ivanov}},\ }\href {\doibase 10.1007/JHEP07(2010)020} {\bibfield  {journal}
  {\bibinfo  {journal} {JHEP}\ }\textbf {\bibinfo {volume} {07}},\ \bibinfo
  {pages} {020} (\bibinfo {year} {2010})},\ \Eprint
  {http://arxiv.org/abs/1004.1802} {arXiv:1004.1802 [hep-th]} \BibitemShut
  {NoStop}%
\bibitem [{\citenamefont {Robens}\ and\ \citenamefont
  {Stefaniak}(2015)}]{Robens:2015gla}%
  \BibitemOpen
  \bibfield  {author} {\bibinfo {author} {\bibfnamefont {T.}~\bibnamefont
  {Robens}}\ and\ \bibinfo {author} {\bibfnamefont {T.}~\bibnamefont
  {Stefaniak}},\ }\href {\doibase 10.1140/epjc/s10052-015-3323-y} {\bibfield
  {journal} {\bibinfo  {journal} {Eur. Phys. J.}\ }\textbf {\bibinfo {volume}
  {C75}},\ \bibinfo {pages} {104} (\bibinfo {year} {2015})},\ \Eprint
  {http://arxiv.org/abs/1501.02234} {arXiv:1501.02234 [hep-ph]} \BibitemShut
  {NoStop}%
\bibitem [{\citenamefont {Muhlleitner}\ \emph {et~al.}(2016)\citenamefont
  {Muhlleitner}, \citenamefont {Sampaio}, \citenamefont {Santos},\ and\
  \citenamefont {Wittbrodt}}]{Muhlleitner:2016mzt}%
  \BibitemOpen
  \bibfield  {author} {\bibinfo {author} {\bibfnamefont {M.}~\bibnamefont
  {Muhlleitner}}, \bibinfo {author} {\bibfnamefont {M.~O.~P.}\ \bibnamefont
  {Sampaio}}, \bibinfo {author} {\bibfnamefont {R.}~\bibnamefont {Santos}}, \
  and\ \bibinfo {author} {\bibfnamefont {J.}~\bibnamefont {Wittbrodt}},\
  }\href@noop {} {\  (\bibinfo {year} {2016})},\ \Eprint
  {http://arxiv.org/abs/1612.01309} {arXiv:1612.01309 [hep-ph]} \BibitemShut
  {NoStop}%
\bibitem [{\citenamefont {Kanemura}\ \emph {et~al.}(1993)\citenamefont
  {Kanemura}, \citenamefont {Kubota},\ and\ \citenamefont
  {Takasugi}}]{Kanemura:1993hm}%
  \BibitemOpen
  \bibfield  {author} {\bibinfo {author} {\bibfnamefont {S.}~\bibnamefont
  {Kanemura}}, \bibinfo {author} {\bibfnamefont {T.}~\bibnamefont {Kubota}}, \
  and\ \bibinfo {author} {\bibfnamefont {E.}~\bibnamefont {Takasugi}},\ }\href
  {\doibase 10.1016/0370-2693(93)91205-2} {\bibfield  {journal} {\bibinfo
  {journal} {Phys. Lett.}\ }\textbf {\bibinfo {volume} {B313}},\ \bibinfo
  {pages} {155} (\bibinfo {year} {1993})},\ \Eprint
  {http://arxiv.org/abs/hep-ph/9303263} {arXiv:hep-ph/9303263 [hep-ph]}
  \BibitemShut {NoStop}%
\bibitem [{\citenamefont {Horejsi}\ and\ \citenamefont
  {Kladiva}(2006)}]{Horejsi:2005da}%
  \BibitemOpen
  \bibfield  {author} {\bibinfo {author} {\bibfnamefont {J.}~\bibnamefont
  {Horejsi}}\ and\ \bibinfo {author} {\bibfnamefont {M.}~\bibnamefont
  {Kladiva}},\ }\href {\doibase 10.1140/epjc/s2006-02472-3} {\bibfield
  {journal} {\bibinfo  {journal} {Eur. Phys. J.}\ }\textbf {\bibinfo {volume}
  {C46}},\ \bibinfo {pages} {81} (\bibinfo {year} {2006})},\ \Eprint
  {http://arxiv.org/abs/hep-ph/0510154} {arXiv:hep-ph/0510154 [hep-ph]}
  \BibitemShut {NoStop}%
\bibitem [{\citenamefont {Chen}\ \emph {et~al.}(2017)\citenamefont {Chen},
  \citenamefont {Kozaczuk},\ and\ \citenamefont {Lewis}}]{Chen:2017qcz}%
  \BibitemOpen
  \bibfield  {author} {\bibinfo {author} {\bibfnamefont {C.-Y.}\ \bibnamefont
  {Chen}}, \bibinfo {author} {\bibfnamefont {J.}~\bibnamefont {Kozaczuk}}, \
  and\ \bibinfo {author} {\bibfnamefont {I.~M.}\ \bibnamefont {Lewis}},\
  }\href@noop {} {\  (\bibinfo {year} {2017})},\ \Eprint
  {http://arxiv.org/abs/1704.05844} {arXiv:1704.05844 [hep-ph]} \BibitemShut
  {NoStop}%
\bibitem [{\citenamefont {Swiezewska}(2015)}]{Swiezewska:2015paa}%
  \BibitemOpen
  \bibfield  {author} {\bibinfo {author} {\bibfnamefont {B.}~\bibnamefont
  {Swiezewska}},\ }\href {\doibase 10.1007/JHEP07(2015)118} {\bibfield
  {journal} {\bibinfo  {journal} {JHEP}\ }\textbf {\bibinfo {volume} {07}},\
  \bibinfo {pages} {118} (\bibinfo {year} {2015})},\ \Eprint
  {http://arxiv.org/abs/1503.07078} {arXiv:1503.07078 [hep-ph]} \BibitemShut
  {NoStop}%
\bibitem [{\citenamefont {Ferreira}\ and\ \citenamefont
  {Swiezewska}(2016)}]{Ferreira:2015pfi}%
  \BibitemOpen
  \bibfield  {author} {\bibinfo {author} {\bibfnamefont {P.~M.}\ \bibnamefont
  {Ferreira}}\ and\ \bibinfo {author} {\bibfnamefont {B.}~\bibnamefont
  {Swiezewska}},\ }\href {\doibase 10.1007/JHEP04(2016)099} {\bibfield
  {journal} {\bibinfo  {journal} {JHEP}\ }\textbf {\bibinfo {volume} {04}},\
  \bibinfo {pages} {099} (\bibinfo {year} {2016})},\ \Eprint
  {http://arxiv.org/abs/1511.02879} {arXiv:1511.02879 [hep-ph]} \BibitemShut
  {NoStop}%
\bibitem [{\citenamefont {Staub}(2008)}]{Staub:2008uz}%
  \BibitemOpen
  \bibfield  {author} {\bibinfo {author} {\bibfnamefont {F.}~\bibnamefont
  {Staub}},\ }\href@noop {} {\  (\bibinfo {year} {2008})},\ \Eprint
  {http://arxiv.org/abs/0806.0538} {arXiv:0806.0538 [hep-ph]} \BibitemShut
  {NoStop}%
\bibitem [{\citenamefont {Staub}(2010)}]{Staub:2009bi}%
  \BibitemOpen
  \bibfield  {author} {\bibinfo {author} {\bibfnamefont {F.}~\bibnamefont
  {Staub}},\ }\href {\doibase 10.1016/j.cpc.2010.01.011} {\bibfield  {journal}
  {\bibinfo  {journal} {Comput.Phys.Commun.}\ }\textbf {\bibinfo {volume}
  {181}},\ \bibinfo {pages} {1077} (\bibinfo {year} {2010})},\ \Eprint
  {http://arxiv.org/abs/0909.2863} {arXiv:0909.2863 [hep-ph]} \BibitemShut
  {NoStop}%
\bibitem [{\citenamefont {Staub}(2011)}]{Staub:2010jh}%
  \BibitemOpen
  \bibfield  {author} {\bibinfo {author} {\bibfnamefont {F.}~\bibnamefont
  {Staub}},\ }\href {\doibase 10.1016/j.cpc.2010.11.030} {\bibfield  {journal}
  {\bibinfo  {journal} {Comput.Phys.Commun.}\ }\textbf {\bibinfo {volume}
  {182}},\ \bibinfo {pages} {808} (\bibinfo {year} {2011})},\ \Eprint
  {http://arxiv.org/abs/1002.0840} {arXiv:1002.0840 [hep-ph]} \BibitemShut
  {NoStop}%
\bibitem [{\citenamefont {Staub}(2013)}]{Staub:2012pb}%
  \BibitemOpen
  \bibfield  {author} {\bibinfo {author} {\bibfnamefont {F.}~\bibnamefont
  {Staub}},\ }\href {\doibase 10.1016/j.cpc.2013.02.019} {\bibfield  {journal}
  {\bibinfo  {journal} {Computer Physics Communications}\ }\textbf {\bibinfo
  {volume} {184}},\ \bibinfo {pages} {pp. 1792} (\bibinfo {year} {2013})},\
  \Eprint {http://arxiv.org/abs/1207.0906} {arXiv:1207.0906 [hep-ph]}
  \BibitemShut {NoStop}%
\bibitem [{\citenamefont {Staub}(2014)}]{Staub:2013tta}%
  \BibitemOpen
  \bibfield  {author} {\bibinfo {author} {\bibfnamefont {F.}~\bibnamefont
  {Staub}},\ }\href {\doibase 10.1016/j.cpc.2014.02.018} {\bibfield  {journal}
  {\bibinfo  {journal} {Comput.Phys.Commun.}\ }\textbf {\bibinfo {volume}
  {185}},\ \bibinfo {pages} {1773} (\bibinfo {year} {2014})},\ \Eprint
  {http://arxiv.org/abs/1309.7223} {arXiv:1309.7223 [hep-ph]} \BibitemShut
  {NoStop}%
\bibitem [{\citenamefont {Staub}(2015)}]{Staub:2015kfa}%
  \BibitemOpen
  \bibfield  {author} {\bibinfo {author} {\bibfnamefont {F.}~\bibnamefont
  {Staub}},\ }\href {\doibase 10.1155/2015/840780} {\bibfield  {journal}
  {\bibinfo  {journal} {Adv. High Energy Phys.}\ }\textbf {\bibinfo {volume}
  {2015}},\ \bibinfo {pages} {840780} (\bibinfo {year} {2015})},\ \Eprint
  {http://arxiv.org/abs/1503.04200} {arXiv:1503.04200 [hep-ph]} \BibitemShut
  {NoStop}%
\bibitem [{\citenamefont {Deshpande}\ and\ \citenamefont
  {Ma}(1978)}]{Deshpande:1977rw}%
  \BibitemOpen
  \bibfield  {author} {\bibinfo {author} {\bibfnamefont {N.~G.}\ \bibnamefont
  {Deshpande}}\ and\ \bibinfo {author} {\bibfnamefont {E.}~\bibnamefont {Ma}},\
  }\href {\doibase 10.1103/PhysRevD.18.2574} {\bibfield  {journal} {\bibinfo
  {journal} {Phys. Rev.}\ }\textbf {\bibinfo {volume} {D18}},\ \bibinfo {pages}
  {2574} (\bibinfo {year} {1978})}\BibitemShut {NoStop}%
\bibitem [{\citenamefont {Barroso}\ \emph {et~al.}(2013)\citenamefont
  {Barroso}, \citenamefont {Ferreira}, \citenamefont {Ivanov},\ and\
  \citenamefont {Santos}}]{Barroso:2013awa}%
  \BibitemOpen
  \bibfield  {author} {\bibinfo {author} {\bibfnamefont {A.}~\bibnamefont
  {Barroso}}, \bibinfo {author} {\bibfnamefont {P.~M.}\ \bibnamefont
  {Ferreira}}, \bibinfo {author} {\bibfnamefont {I.~P.}\ \bibnamefont
  {Ivanov}}, \ and\ \bibinfo {author} {\bibfnamefont {R.}~\bibnamefont
  {Santos}},\ }\href {\doibase 10.1007/JHEP06(2013)045} {\bibfield  {journal}
  {\bibinfo  {journal} {JHEP}\ }\textbf {\bibinfo {volume} {06}},\ \bibinfo
  {pages} {045} (\bibinfo {year} {2013})},\ \Eprint
  {http://arxiv.org/abs/1303.5098} {arXiv:1303.5098 [hep-ph]} \BibitemShut
  {NoStop}%
\bibitem [{\citenamefont {Coleman}\ and\ \citenamefont
  {Weinberg}(1973)}]{Coleman:1973jx}%
  \BibitemOpen
  \bibfield  {author} {\bibinfo {author} {\bibfnamefont {S.~R.}\ \bibnamefont
  {Coleman}}\ and\ \bibinfo {author} {\bibfnamefont {E.~J.}\ \bibnamefont
  {Weinberg}},\ }\href {\doibase 10.1103/PhysRevD.7.1888} {\bibfield  {journal}
  {\bibinfo  {journal} {Phys. Rev.}\ }\textbf {\bibinfo {volume} {D7}},\
  \bibinfo {pages} {1888} (\bibinfo {year} {1973})}\BibitemShut {NoStop}%
\bibitem [{\citenamefont {Machacek}\ and\ \citenamefont
  {Vaughn}(1983)}]{Machacek:1983tz}%
  \BibitemOpen
  \bibfield  {author} {\bibinfo {author} {\bibfnamefont {M.~E.}\ \bibnamefont
  {Machacek}}\ and\ \bibinfo {author} {\bibfnamefont {M.~T.}\ \bibnamefont
  {Vaughn}},\ }\href {\doibase 10.1016/0550-3213(83)90610-7} {\bibfield
  {journal} {\bibinfo  {journal} {Nucl. Phys.}\ }\textbf {\bibinfo {volume}
  {B222}},\ \bibinfo {pages} {83} (\bibinfo {year} {1983})}\BibitemShut
  {NoStop}%
\bibitem [{\citenamefont {Machacek}\ and\ \citenamefont
  {Vaughn}(1984)}]{Machacek:1983fi}%
  \BibitemOpen
  \bibfield  {author} {\bibinfo {author} {\bibfnamefont {M.~E.}\ \bibnamefont
  {Machacek}}\ and\ \bibinfo {author} {\bibfnamefont {M.~T.}\ \bibnamefont
  {Vaughn}},\ }\href {\doibase 10.1016/0550-3213(84)90533-9} {\bibfield
  {journal} {\bibinfo  {journal} {Nucl. Phys.}\ }\textbf {\bibinfo {volume}
  {B236}},\ \bibinfo {pages} {221} (\bibinfo {year} {1984})}\BibitemShut
  {NoStop}%
\bibitem [{\citenamefont {Machacek}\ and\ \citenamefont
  {Vaughn}(1985)}]{Machacek:1984zw}%
  \BibitemOpen
  \bibfield  {author} {\bibinfo {author} {\bibfnamefont {M.~E.}\ \bibnamefont
  {Machacek}}\ and\ \bibinfo {author} {\bibfnamefont {M.~T.}\ \bibnamefont
  {Vaughn}},\ }\href {\doibase 10.1016/0550-3213(85)90040-9} {\bibfield
  {journal} {\bibinfo  {journal} {Nucl. Phys.}\ }\textbf {\bibinfo {volume}
  {B249}},\ \bibinfo {pages} {70} (\bibinfo {year} {1985})}\BibitemShut
  {NoStop}%
\bibitem [{\citenamefont {Camargo-Molina}\ \emph
  {et~al.}(2013{\natexlab{d}})\citenamefont {Camargo-Molina}, \citenamefont
  {O'Leary}, \citenamefont {Porod},\ and\ \citenamefont
  {Staub}}]{Camargo-Molina:2013qva}%
  \BibitemOpen
  \bibfield  {author} {\bibinfo {author} {\bibfnamefont {J.~E.}\ \bibnamefont
  {Camargo-Molina}}, \bibinfo {author} {\bibfnamefont {B.}~\bibnamefont
  {O'Leary}}, \bibinfo {author} {\bibfnamefont {W.}~\bibnamefont {Porod}}, \
  and\ \bibinfo {author} {\bibfnamefont {F.}~\bibnamefont {Staub}},\ }\href
  {\doibase 10.1140/epjc/s10052-013-2588-2} {\bibfield  {journal} {\bibinfo
  {journal} {Eur. Phys. J.}\ }\textbf {\bibinfo {volume} {C73}},\ \bibinfo
  {pages} {2588} (\bibinfo {year} {2013}{\natexlab{d}})},\ \Eprint
  {http://arxiv.org/abs/1307.1477} {arXiv:1307.1477 [hep-ph]} \BibitemShut
  {NoStop}%
\bibitem [{\citenamefont {Porod}(2003)}]{Porod:2003um}%
  \BibitemOpen
  \bibfield  {author} {\bibinfo {author} {\bibfnamefont {W.}~\bibnamefont
  {Porod}},\ }\href {\doibase 10.1016/S0010-4655(03)00222-4} {\bibfield
  {journal} {\bibinfo  {journal} {Comput.Phys.Commun.}\ }\textbf {\bibinfo
  {volume} {153}},\ \bibinfo {pages} {275} (\bibinfo {year} {2003})},\ \Eprint
  {http://arxiv.org/abs/hep-ph/0301101} {arXiv:hep-ph/0301101 [hep-ph]}
  \BibitemShut {NoStop}%
\bibitem [{\citenamefont {Porod}\ and\ \citenamefont
  {Staub}(2012)}]{Porod:2011nf}%
  \BibitemOpen
  \bibfield  {author} {\bibinfo {author} {\bibfnamefont {W.}~\bibnamefont
  {Porod}}\ and\ \bibinfo {author} {\bibfnamefont {F.}~\bibnamefont {Staub}},\
  }\href {\doibase 10.1016/j.cpc.2012.05.021} {\bibfield  {journal} {\bibinfo
  {journal} {Comput.Phys.Commun.}\ }\textbf {\bibinfo {volume} {183}},\
  \bibinfo {pages} {2458} (\bibinfo {year} {2012})},\ \Eprint
  {http://arxiv.org/abs/1104.1573} {arXiv:1104.1573 [hep-ph]} \BibitemShut
  {NoStop}%
\bibitem [{\citenamefont {Bechtle}\ \emph {et~al.}(2010)\citenamefont
  {Bechtle}, \citenamefont {Brein}, \citenamefont {Heinemeyer}, \citenamefont
  {Weiglein},\ and\ \citenamefont {Williams}}]{Bechtle:2008jh}%
  \BibitemOpen
  \bibfield  {author} {\bibinfo {author} {\bibfnamefont {P.}~\bibnamefont
  {Bechtle}}, \bibinfo {author} {\bibfnamefont {O.}~\bibnamefont {Brein}},
  \bibinfo {author} {\bibfnamefont {S.}~\bibnamefont {Heinemeyer}}, \bibinfo
  {author} {\bibfnamefont {G.}~\bibnamefont {Weiglein}}, \ and\ \bibinfo
  {author} {\bibfnamefont {K.~E.}\ \bibnamefont {Williams}},\ }\href {\doibase
  10.1016/j.cpc.2009.09.003} {\bibfield  {journal} {\bibinfo  {journal}
  {Comput. Phys. Commun.}\ }\textbf {\bibinfo {volume} {181}},\ \bibinfo
  {pages} {138} (\bibinfo {year} {2010})},\ \Eprint
  {http://arxiv.org/abs/0811.4169} {arXiv:0811.4169 [hep-ph]} \BibitemShut
  {NoStop}%
\bibitem [{\citenamefont {Bechtle}\ \emph {et~al.}(2014)\citenamefont
  {Bechtle}, \citenamefont {Brein}, \citenamefont {Heinemeyer}, \citenamefont
  {Stål}, \citenamefont {Stefaniak}, \citenamefont {Weiglein},\ and\
  \citenamefont {Williams}}]{Bechtle:2013wla}%
  \BibitemOpen
  \bibfield  {author} {\bibinfo {author} {\bibfnamefont {P.}~\bibnamefont
  {Bechtle}}, \bibinfo {author} {\bibfnamefont {O.}~\bibnamefont {Brein}},
  \bibinfo {author} {\bibfnamefont {S.}~\bibnamefont {Heinemeyer}}, \bibinfo
  {author} {\bibfnamefont {O.}~\bibnamefont {Stål}}, \bibinfo {author}
  {\bibfnamefont {T.}~\bibnamefont {Stefaniak}}, \bibinfo {author}
  {\bibfnamefont {G.}~\bibnamefont {Weiglein}}, \ and\ \bibinfo {author}
  {\bibfnamefont {K.~E.}\ \bibnamefont {Williams}},\ }\href {\doibase
  10.1140/epjc/s10052-013-2693-2} {\bibfield  {journal} {\bibinfo  {journal}
  {Eur. Phys. J.}\ }\textbf {\bibinfo {volume} {C74}},\ \bibinfo {pages} {2693}
  (\bibinfo {year} {2014})},\ \Eprint {http://arxiv.org/abs/1311.0055}
  {arXiv:1311.0055 [hep-ph]} \BibitemShut {NoStop}%
\bibitem [{\citenamefont {Nierste}\ and\ \citenamefont
  {Riesselmann}(1996)}]{Nierste:1995zx}%
  \BibitemOpen
  \bibfield  {author} {\bibinfo {author} {\bibfnamefont {U.}~\bibnamefont
  {Nierste}}\ and\ \bibinfo {author} {\bibfnamefont {K.}~\bibnamefont
  {Riesselmann}},\ }\href {\doibase 10.1103/PhysRevD.53.6638} {\bibfield
  {journal} {\bibinfo  {journal} {Phys. Rev.}\ }\textbf {\bibinfo {volume}
  {D53}},\ \bibinfo {pages} {6638} (\bibinfo {year} {1996})},\ \Eprint
  {http://arxiv.org/abs/hep-ph/9511407} {arXiv:hep-ph/9511407 [hep-ph]}
  \BibitemShut {NoStop}%
\bibitem [{\citenamefont {Wainwright}(2012)}]{Wainwright:2011kj}%
  \BibitemOpen
  \bibfield  {author} {\bibinfo {author} {\bibfnamefont {C.~L.}\ \bibnamefont
  {Wainwright}},\ }\href {\doibase 10.1016/j.cpc.2012.04.004} {\bibfield
  {journal} {\bibinfo  {journal} {Comput. Phys. Commun.}\ }\textbf {\bibinfo
  {volume} {183}},\ \bibinfo {pages} {2006} (\bibinfo {year} {2012})},\ \Eprint
  {http://arxiv.org/abs/1109.4189} {arXiv:1109.4189 [hep-ph]} \BibitemShut
  {NoStop}%
\end{thebibliography}%

\end{document}